\documentclass[journal]{IEEEtran}

\usepackage{cite}
\usepackage{pifont}
\usepackage[table]{xcolor}
\usepackage{graphicx}
\usepackage{booktabs}
\usepackage{tabularx}
\usepackage{array}
\usepackage{amsmath}
\usepackage{amssymb}
\usepackage{tikz}
\usetikzlibrary{arrows.meta,positioning,shapes.geometric,fit,decorations.pathmorphing}
\usepackage{url}
\usepackage{flushend} 

\newcommand{\yes}{\textcolor{teal}{\ding{108}}}
\newcommand{\maybe}{\textcolor{teal}{\ding{109}}}
\newcommand{\no}{\textcolor{red}{\ding{55}}}
\newcommand{\vP}{\mathbf{v}}
\newcolumntype{Y}{>{\raggedright\arraybackslash}X}
\newcolumntype{C}[1]{>{\centering\arraybackslash}p{#1}}

\begin{document}
\bstctlcite{IEEEexample:BSTcontrol}

\title{Automotive Hardware Attacks: An Architect's Guide to TARA}

\author{Jakub Breier and Xiaolu Hou
\thanks{J. Breier is with TTControl GmbH, Vienna, Austria, e-mail: jbreier@jbreier.com\\
X. Hou is with Slovak University of Technology, Bratislava, Slovakia, e-mail: houxiaolu.email@gmail.com}}

\maketitle

\begin{abstract}
Threat analysis and risk assessment (TARA) according to ISO/SAE 21434 treats implementation-level hardware attacks inconsistently. We show that two of the three attack-feasibility approaches exemplified in the standard rate every attack path that requires physical access as very low by construction, independent of the implementation, while the third can assign the same path any of the four ratings, depending on the assumed state of attacker knowledge and on the attacker profile. This paper presents a hardware-aware extension of the TARA that covers side-channel analysis (SCA), fault-injection attacks (FIA), and abuse of debug and test interfaces. It adds an attacker model for physical access over the vehicle lifecycle; a path-dominance relation on the attack-potential factors that identifies when a simpler path makes a sophisticated one irrelevant; a three-level hardware relevance gate (H0 to H2) with an explicit decision rule that fixes whether resistance of a hardware step may be argued, must be documented, or must be demonstrated by test; and a rating rule under which a hardware step receives no credit for implementation-specific resistance until the required evidence exists. The gate is positioned relative to cybersecurity assurance levels and targeted attack feasibility. The method is applied to a reference gateway architecture, and its structural inputs are examined on ten published attacks on automotive components. In seven of the ten studies, the demonstrated attack included a debug, programming, boot, diagnostic, or update mechanism, whereas only four studies included an attack on a cryptographic computation. 
\end{abstract}


\section{Introduction}
\IEEEPARstart{V}{ehicle} electrical and electronic architectures are moving from many function-specific electronic control units (ECUs) toward domain controllers and central vehicle computers with permanent external connectivity~\cite{bandur2021centralized,montanaro2019towards}. After experimental analyses and remote compromises of production vehicles~\cite{koscher2010experimental,miller2015remote}, cybersecurity became a lifecycle engineering discipline: ISO/SAE 21434 defines the engineering framework, and UN Regulation No.~155 makes a cybersecurity management system and a vehicle-level risk assessment a condition for type approval~\cite{iso21434,r155}. Much of the protection developed since then operates at the network level. Authentication and key-agreement schemes for in-vehicle buses, including~\cite{groza2020cmac,musuroi2021groupkey,murvay2018j1939}, and AUTOSAR Secure Onboard Communication (SecOC)~\cite{autosar_secoc} assume that each ECU keeps its keys confidential and that its boot and update decisions cannot be bypassed.

Threat analysis and risk assessment (TARA) connects the architecture to cybersecurity risk. Assets and damage scenarios are identified, threat scenarios are derived, attack paths are analyzed, attack feasibility and impact are rated, and risk-treatment decisions lead to cybersecurity goals and requirements. Automotive TARA methods and extensions include EVITA~\cite{ruddle2009evita}, HEAVENS~2.0~\cite{lautenbach2021heavens}, ThreatSurf~\cite{zelle2022threatsurf}, STRIDE-based frameworks~\cite{abuabed2023stride}, and TARA~2.0 for connected and automated vehicles~\cite{benyahya2025tara}; a survey is given in~\cite{luo2021tara}. These methods operate mainly at the architecture, software, and network levels.

Implementation-level hardware attacks target exactly the assumptions named above. Side-channel analysis (SCA) recovers secrets from physical leakage such as power consumption or electromagnetic emanation. Fault-injection attacks (FIA) perturb a computation to alter data or a security decision~\cite{hou2024cryptography,breier2022practical}. Debug and test interfaces such as JTAG, SWD, or BDM give direct access to memory and execution when production access control is missing or can be defeated. All three have been demonstrated on automotive components: key recovery from keyless-entry and immobilizer systems~\cite{eisenbarth2008keeloq,garcia2016lock,wouters2020dst80}, SCA and FIA against authenticated in-vehicle communication~\cite{oflynn2018securecan,ebina2023secoc}, fault-assisted firmware extraction and bootloader bypass on production ECUs~\cite{milburn2018glitches,oflynn2020bam,pozzobon2023fuzzy}, and compromise of all computing subsystems of a connected electric vehicle~\cite{kuhnapfel2025three}. The topic has also reached standardization and regulation. SAE J3101-4, published in 2026, surveys SCA and FIA with an impact on automotive embedded systems and lists countermeasures that manufacturers can consider~\cite{saej3101_4}. Annex~5 of UN Regulation No.~155 lists the extraction of cryptographic keys, remainders from development such as debug and JTAG ports, and the manipulation of electronic hardware among the threats and vulnerabilities that the manufacturer's risk assessment shall consider~\cite{r155}.

Despite this, the TARA process gives little support for two questions. The first is how a hardware-assisted attack path should be rated. Section~\ref{sec:analysis} shows that the answer depends on the rating approach to a degree that makes the result arbitrary: two of the three approaches exemplified in ISO/SAE 21434 rate every path that requires physical access as very low, and the third can assign the same path any of the four ratings. The second is when such a path justifies dedicated hardware-security evidence or testing. Testing every ECU against every physical attack is not economically realistic. Excluding all of them because they need physical access contradicts both the published attacks and the regulation.

This paper addresses both questions at the methodology level. Its contributions can be summarized as follows:
\begin{itemize}
    \item an analysis of how the three example approaches of ISO/SAE 21434, based on the attack vector, the Common Vulnerability Scoring System (CVSS), and the attack potential, rate implementation-level attack paths, showing a structural cap in the first two and a high sensitivity of the third to inputs that are unrelated to the implementation (Section~\ref{sec:analysis});
    \item an attacker model for physical access over the vehicle lifecycle and a modeling rule that keeps SCA, FIA, and debug-interface abuse at the level of attack methods within attack paths (Sections~\ref{sec:background} and~\ref{sec:method});
    \item a path-dominance relation on the attack-potential factors, which identifies when a simpler path makes a specialist SCA or FIA evaluation irrelevant (Section~\ref{sec:paths});
    \item a three-level hardware relevance gate (H0 to H2) with an explicit decision rule, baseline checks, and evidence classes; a rating rule under which a hardware step receives no credit for implementation-specific resistance until the required evidence exists; and the positioning of both against cybersecurity assurance levels (CAL) and targeted attack feasibility (TAF) (Sections~\ref{sec:gate} to~\ref{sec:cal});
    \item an application to a reference gateway architecture and a retrospective structural analysis of ten published attacks on automotive components (Sections~\ref{sec:example} and~\ref{sec:retro}).
\end{itemize}

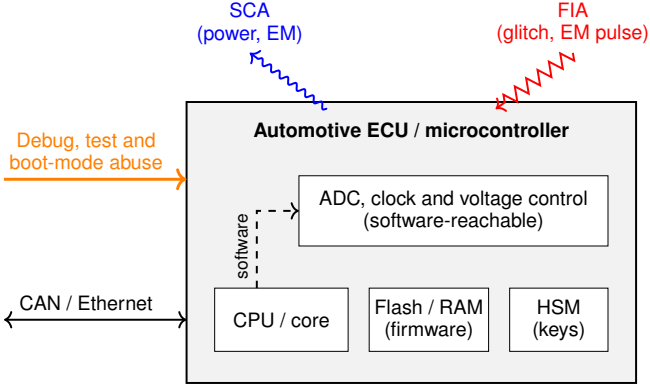
\begin{figure}[t]
    \centering
    \resizebox{0.98\linewidth}{!}{%
    \begin{tikzpicture}[font=\sffamily\footnotesize]
        \draw[thick, fill=gray!10] (0,0) rectangle (6.4,4);
        \node at (3.2,3.6) {\textbf{Automotive ECU / microcontroller}};

        \draw[fill=white] (0.4,0.45) rectangle (2.3,1.35) node[midway] {CPU / core};
        \draw[fill=white] (2.6,0.45) rectangle (4.3,1.35) node[midway, align=center] {Flash / RAM\\(firmware)};
        \draw[fill=white] (4.6,0.45) rectangle (6.0,1.35) node[midway, align=center] {HSM\\(keys)};
        \draw[fill=white] (1.6,1.95) rectangle (6.0,2.95) node[midway, align=center] {ADC, clock and voltage control\\(software-reachable)};

        \draw[->, thick, dashed] (1.0,1.35) -- (1.0,2.45) -- (1.6,2.45);
        \node[rotate=90, font=\sffamily\scriptsize] at (0.78,1.95) {software};

        \draw[<->, thick] (-2.6,0.9) -- (0,0.9) node[pos=0.45, above] {CAN / Ethernet};
        \draw[->, very thick, orange] (-2.6,2.9) -- (0,2.9) node[pos=0.45, above, align=center] {Debug, test and\\boot-mode abuse};

        \draw[blue, thick, decorate, decoration={snake, amplitude=.4mm, segment length=1.5mm, post length=1mm}, ->]
            (2,3.9) -- (0.9,4.7) node[pos=1, above, align=center] {SCA\\(power, EM)};

        \draw[red, thick, decorate, decoration={zigzag, amplitude=1mm, segment length=2mm, post length=1.5mm}, ->]
            (5.5,4.7) -- (4.4,3.9) node[pos=0, above, align=center] {FIA\\(glitch, EM pulse)};
    \end{tikzpicture}}
    \caption{Implementation-level attack classes considered in this paper. SCA observes information leakage, FIA perturbs computation or storage, and abuse of debug, test, and boot-mode interfaces gives direct access to memory and execution. In software-mediated variants (dashed), compromised software uses on-chip measurement or regulation resources instead of external equipment.}
    \label{fig:ecu}
\end{figure}

\section{Background and Scope}
\label{sec:background}
\subsection{TARA in ISO/SAE 21434}
ISO/SAE 21434 specifies the TARA as a set of activities: asset identification, threat scenario identification, impact rating, attack path analysis, attack feasibility rating, risk value determination, and risk treatment decision~\cite{iso21434}. Damage scenarios describe adverse consequences for road users and are rated in the categories safety, financial, operational, and privacy (S, F, O, P) on four levels from negligible to severe. Threat scenarios describe how a cybersecurity property of an asset is compromised. Each attack path that realizes a threat scenario receives an attack feasibility rating on four levels from very low to high, and the risk value, an integer from 1 to 5, is determined from impact and feasibility, for example with a risk matrix. If several paths realize the same threat scenario, the standard gives as an example assigning the maximum of their feasibility ratings. Risk is then avoided, reduced, shared, or retained. When a risk is shared, or retained on the basis of assumptions, the rationale is recorded as a cybersecurity claim, which can be subject to monitoring.

The standard does not prescribe how feasibility is rated. Its Annex~G gives three example approaches. The attack-potential-based approach, derived from ISO/IEC 18045~\cite{iso18045}, sums ratings for elapsed time, specialist expertise, knowledge of the item or component, window of opportunity, and equipment. The CVSS-based approach uses the exploitability metrics of CVSS\footnote{The CVSS version used in this paper is~v3.1~\cite{cvss31}.}. The attack-vector-based approach uses only the attack vector (network, adjacent, local, or physical). Section~\ref{sec:analysis} analyzes how these approaches treat physical paths.

\subsection{Implementation-Level Attack Classes}
Fig.~\ref{fig:ecu} shows the attack classes considered.

\emph{Side-channel analysis} is passive: the adversary infers sensitive information from implementation-dependent leakage such as timing, power consumption, or electromagnetic emanation~\cite{kocher1999differential,hou2024cryptography}. Key recovery is the most common use, but side channels can also expose control flow and the parameters of machine-learning models~\cite{eisenbarth2010building,batina2019csi}. Countermeasures include masking, hiding, and hardened cryptographic accelerators or hardware security modules (HSMs)~\cite{prouff2013masking,nikova2006threshold,hettwer2020lightweight}, complemented by architectural measures such as key diversification and limits on the use of long-lived secrets.

\emph{Fault injection} is active: the adversary perturbs the device, for example with voltage or clock glitches, electromagnetic pulses, or lasers, to alter data, control flow, or a security decision~\cite{baksi2023survey}. Equipment is available from tens of US dollars for voltage glitching and from a few hundred US dollars for electromagnetic pulse injection~\cite{breier2022practical}. The mechanisms required for functional safety do not provide resistance by themselves: lockstep microcontrollers certified to Automotive Safety Integrity Level D have been shown to be susceptible~\cite{wiersma2017safety}. On a microcontroller, a fault can make the processor skip individual instructions~\cite{breier2015laser}. A single successful fault is therefore sufficient when it hits a decision on which other controls depend, such as secure boot, update authentication, debug authorization, or a diagnostic access check.

\emph{Debug and test interfaces} (JTAG, SWD, BDM, boot-mode loaders, serial consoles) are needed in development, manufacturing, and service. If they remain accessible in production, they allow firmware extraction, memory access, and execution control with standard tools~\cite{zhang2021csef,herrewegen2018bonnet,garcia2016lock}. Read-out protection and boot-mode passwords are themselves decisions that fault injection can defeat~\cite{bozzato2019shaping,herrewegen2021boots,oflynn2020bam,wouters2020dst80}. Debug access is therefore treated as an attack method in its own right and not only as a precondition for SCA or FIA.

\emph{Software-mediated variants} do not require physical access. Analog-to-digital converters can give software an observation that is correlated with computation on the same chip~\cite{gnad2019leaky}, software-controlled dynamic voltage and frequency scaling (DVFS feature) can induce faults~\cite{tang2017clkscrew,murdock2020plundervolt}, and memory-disturbance effects can corrupt adjacent cells~\cite{mutlu2017rowhammer} (Rowhammer). These effects have mostly been shown on platforms other than automotive microcontrollers. Their relevance for an ECU depends on whether such resources exist and can be reached by compromised software.

\subsection{Attacker Profiles Over the Vehicle Lifecycle}
\label{sec:profiles}
Whether a hardware-assisted path is credible depends on who has access to the component, for how long, and whether the attacked unit has to remain functional. Table~\ref{tab:profiles} lists the five profiles used in this paper, together with the window-of-opportunity level that we assign to each.

\begin{table}[t]
\centering
\caption{Attacker profiles for hardware-assisted attack paths and the window-of-opportunity level assigned to each.}
\label{tab:profiles}
\scriptsize
\begin{tabularx}{\columnwidth}{p{2.35cm}Yp{1.2cm}}
\toprule
\textbf{Profile} & \textbf{Access to the component} & \textbf{Window} \\
\midrule
A1 Owner or keeper & Own vehicle without time limit; can buy spare units & Unlimited \\
A2 Workshop & Authorized access for hours to days, tools available & Moderate \\
A3 Unauthorized third party & Victim vehicle for minutes to hours; unnoticed removal and reinstallation of an ECU is rarely practical & Difficult \\
A4 Analyst with donor unit & Unit of the same type from the parts market, in a laboratory, without time limit & Unlimited \\
A5 Remote, with code execution & No physical access; only resources that software can reach & By network reachability \\
\bottomrule
\end{tabularx}
\end{table}

The owner profile is easily overlooked. ISO/SAE 21434 rates impact for road users, so an owner who attacks their own vehicle seems out of scope. This is not the case when the damage affects others, as with a defeated speed limiter, tampered emission controls or odometer, or a disabled safety monitor, or when the organization adds impact categories for its own assets, which the standard allows. The motives of owners and of other stakeholders with physical access are discussed for a production vehicle in~\cite{kuhnapfel2025three}. For the owner the window of opportunity is unlimited by definition. The donor-unit profile has the same property for every asset that is identical across units, because any unit of the type serves the attacker.

\subsection{Scope}
Hardware Trojans, counterfeit or substituted components, and malicious modification during fabrication or assembly are relevant automotive threats~\cite{bozdal2018trojan}. They differ from the three classes above in that the adversarial action is introduced in the supply chain and not applied to a legitimately manufactured ECU. Trust boundary, attacker opportunity, and evidence sources differ accordingly. These threats remain branches of the broader TARA wherever the item definition makes them credible, but they are outside the gate proposed here, which decides the assurance depth for attack paths against a legitimately manufactured ECU.

\section{Related Work}
\subsection{Automotive TARA Methods}
Early work established the need for a TARA in analogy to the hazard analysis of functional safety~\cite{ward2013tara}. EVITA combined attack trees with an attack-potential rating adapted from ISO/IEC 18045~\cite{ruddle2009evita}. HEAVENS~2.0 added damage scenarios and attack-path analysis for alignment with ISO/SAE 21434~\cite{lautenbach2021heavens}. ThreatSurf automates attack-surface assessment with reusable attack building blocks~\cite{zelle2022threatsurf}, Abuabed \emph{et al.} combine STRIDE, attack trees, and exploitability scoring~\cite{abuabed2023stride}, TARA~2.0 extends the analysis to connected and automated vehicles~\cite{benyahya2025tara}, and tool support for asset-driven analysis exists~\cite{schmittner2021threatget}. The survey in~\cite{luo2021tara} concentrates on network and software threats. Where hardware attacks appear, they are generic building blocks with fixed ratings. ThreatSurf, for example, rates the extraction of firmware or of keys in a hardware security module by a hardware attack as requiring bespoke equipment and a difficult window of opportunity, which yields a very low feasibility independent of the implementation~\cite{zelle2022threatsurf}.

\subsection{Attack Feasibility Rating}
The attack-potential approach has been criticized from within the industry. Greiner \emph{et al.} report that it leads to inconsistent results that often do not reflect reality, single out the window of opportunity as a factor that analysts understand very differently, and propose modified scales~\cite{greiner2022supplier}. Kamtam \emph{et al.} find that the window-of-opportunity parameter does not differentiate remote from physical attacks and estimate it by simulation instead of expert judgment~\cite{kamtam2025wolves}. For machine-learning models in automated driving, Schr\"oder and Breier measure the extent of damage and the attacker's effort with separate indicators and conclude that the effort cannot be expressed as a single value~\cite{schroder2024rmf}. In the certification of smart cards and similar devices, the Joint Interpretation Library refines ISO/IEC 18045 by rating the \emph{identification} of an attack, that is, the effort to create it and to show that it works, separately from its \emph{exploitation} on another instance once the method is known~\cite{jil2024attackpotential}. ISO/SAE 21434 does not make this distinction. None of these works examines how the three approaches of Annex~G treat implementation-level paths.

\subsection{Assurance Levels}
Annex~E of ISO/SAE 21434 introduces cybersecurity assurance levels (CAL1 to CAL4) that scale the rigor of cybersecurity activities; its example determines the CAL from the impact and the attack vector. ISO/SAE PAS 8475, under publication at the time of writing, develops the CAL concept and introduces targeted attack feasibility (TAF), the attack feasibility to be achieved after cybersecurity controls are applied~\cite{isopas8475}. SAE J3101-4 offers the supply chain a common language for communicating SCA and FIA mitigation requirements~\cite{saej3101_4}. Section~\ref{sec:cal} relates the proposed gate to CAL and TAF.

\subsection{Hardware Attacks on Automotive Components}
Published attacks cover keyless entry and immobilizers, in-vehicle communication, firmware extraction and bootloaders, and vehicle computers; Section~\ref{sec:retro} analyzes ten of them. These studies demonstrate that individual paths are feasible. They do not translate the paths into damage scenarios, competing-path reasoning, or decisions on supplier evidence and testing. Generic TARA methods, in turn, explain how risk is assessed but not when an implementation-level path deserves an explicit branch and dedicated evidence. The method proposed here addresses this interface.

\section{Physical Attack Paths Under the Annex G Approaches}
\label{sec:analysis}
This section examines how each of the three example approaches rates an attack path that contains a hardware step. We use the example values of Annex~G, the example risk matrix of Annex~H, and the example CAL determination of Annex~E of the ISO/SAE 21434:2021~\cite{iso21434}. 

\subsection{Attack-Vector-Based and CVSS-Based Approaches}
The attack-vector-based approach maps the vector directly to the rating: network to high, adjacent to medium, local to low, and physical to very low. A path that needs physical access is therefore rated very low whatever the asset, the attacker, or the implementation.

The CVSS-based approach computes the exploitability
\begin{equation}
E = 8.22 \cdot V \cdot C \cdot P \cdot U
\label{eq:cvss}
\end{equation}
from the CVSS metrics attack vector $V$, attack complexity $C$, privileges required $P$, and user interaction $U$, and maps $E$ to the four ratings by thresholds, with $E \leq 1.05$ rated very low. With the weights of CVSS v3.1~\cite{cvss31}, $V = 0.20$ for a physical vector and the product of the other three metrics is at most $0.77 \cdot 0.85 \cdot 0.85 = 0.556$, so that
\begin{equation}
E \leq 8.22 \cdot 0.20 \cdot 0.556 = 0.91
\end{equation}
for every path with a physical vector. A typical hardware path with high complexity, no privileges, and no user interaction gives $E = 0.52$. The result is again very low in all cases. For comparison, the maximum of $E$ is 2.52 for a local, 2.84 for an adjacent, and 3.89 for a network vector.

With the example risk matrix, a very low feasibility gives a risk value of 1 for moderate and major impact and of 2 for severe impact. Under both approaches, an open debug port in front of a fleet-wide key and a key inside an evaluated secure element receive the same rating and the same risk value. The CAL example of Annex~E has the same structure: for a physical vector it assigns CAL1 for moderate and major impact and CAL2 for severe impact, the lowest level in each row.

\subsection{Attack-Potential-Based Approach}
Table~\ref{tab:apvalues} lists the example values. The sum is not capped for physical paths, so this approach can express both a trivial debug read-out and a hardened implementation. Its result for hardware steps is nevertheless dominated by two inputs that are unrelated to the implementation under analysis.

\begin{table}[t]
\centering
\caption{Example values of the attack-potential-based approach in ISO/SAE 21434, Annex~G~\cite{iso21434}.}
\label{tab:apvalues}
\scriptsize
\begin{tabularx}{\columnwidth}{p{2.4cm}Y}
\toprule
\textbf{Factor} & \textbf{Levels (value)} \\
\midrule
Elapsed time (ET) & $\leq$1 day (0), $\leq$1 week (1), $\leq$1 month (4), $\leq$6 months (17), $>$6 months (19) \\
Specialist expertise (SE) & layman (0), proficient (3), expert (6), multiple experts (8) \\
Knowledge of the item (KN) & public (0), restricted (3), confidential (7), strictly confidential (11) \\
Window of opportunity (WO) & unlimited (0), easy (1), moderate (4), difficult (10) \\
Equipment (EQ) & standard (0), specialized (4), bespoke (7), multiple bespoke (9) \\
\midrule
Rating by sum & 0--13 high (H), 14--19 medium (M), 20--24 low (L), $\geq$25 very low (VL) \\
\bottomrule
\end{tabularx}
\end{table}

The first input is the state of knowledge. Elapsed time covers identifying a vulnerability and developing and applying an exploit, and it is rated against the state of expert knowledge at the time of rating; the required expertise depends on the same state. For a hardware step this state changes abruptly when an attack on the component family is published: the first attacker needs weeks or months of characterization, whereas a follower needs the published procedure and the same equipment. The example scale amplifies the change, since the value for elapsed time rises from 4 for up to one month to 17 for up to six months.

The second input is the attacker profile, which sets the window of opportunity. The same step is rated 0 for an owner or for an analyst with a donor unit and 10 for a third party who would have to remove and reinstall an ECU unnoticed.

Table~\ref{tab:sens} quantifies both effects for four hardware steps against implementations without specific countermeasures. The factor levels are reference values that we derived from published attacks: read-out with standard programming tools~\cite{garcia2016lock}; voltage glitching of an authorization check, which took about three days for a complete firmware image in~\cite{milburn2018glitches} and has been shown with low-cost equipment~\cite{bozzato2019shaping}; correlation power analysis of an unprotected cipher~\cite{oflynn2018securecan,eisenbarth2008keeloq}; and electromagnetic fault injection on a boot or update verification, where position and timing have to be searched again for each target~\cite{oflynn2020bam,pozzobon2023fuzzy}. Debug probes are rated as standard equipment. Oscilloscopes, glitchers, and electromagnetic pulse generators are rated as specialized equipment, in line with ISO/IEC 18045, which names power analysis tools as an example of that category~\cite{iso18045}, and with their cost~\cite{breier2022practical}. Knowledge is rated as public throughout, because the cited attacks were carried out by third parties without confidential information.

\begin{table*}[t]
\centering
\caption{Attack feasibility of four hardware steps against implementations without specific countermeasures under the three approaches of Annex~G. For the attack-potential-based approach, each entry gives the sum and the rating; the window of opportunity (WO) is added to the factor levels in the second and third column. The follower columns are the reference values of the no-credit rating (rule~3 in Section~\ref{sec:rating}).}
\label{tab:sens}
\scriptsize
\begin{tabularx}{\textwidth}{Y C{1.7cm} C{1.7cm} C{1.25cm} C{1.25cm} C{1.2cm} C{1.2cm} C{1.2cm} C{1.2cm}}
\toprule
 & \multicolumn{2}{c}{\textbf{Levels (ET, SE, KN, EQ)}} & & & \multicolumn{4}{c}{\textbf{Attack-potential-based: sum (rating)}} \\
\cmidrule(lr){2-3}\cmidrule(lr){6-9}
\textbf{Hardware step} & First attacker & Follower & \textbf{Attack-vector-based} & \textbf{CVSS-based ($E$)} & First attacker, WO\,=\,0 & Follower, A1/A4, WO\,=\,0 & Follower, A2, WO\,=\,4 & Follower, A3, WO\,=\,10 \\
\midrule
Read-out through an open debug or programming interface & (1, 3, 0, 0) & (0, 3, 0, 0) & VL & VL (0.91) & 4 (H) & 3 (H) & 7 (H) & 13 (H) \\
\addlinespace
Voltage glitch on a read-out, boot, or diagnostic authorization check & (4, 6, 0, 4) & (1, 3, 0, 4) & VL & VL (0.52) & 14 (M) & 8 (H) & 12 (H) & 18 (M) \\
\addlinespace
Correlation power analysis of an unprotected cipher & (4, 6, 0, 4) & (1, 3, 0, 4) & VL & VL (0.52) & 14 (M) & 8 (H) & 12 (H) & 18 (M) \\
\addlinespace
Electromagnetic fault on a boot or update verification & (17, 6, 0, 4) & (1, 6, 0, 4) & VL & VL (0.52) & 27 (VL) & 11 (H) & 15 (M) & 21 (L) \\
\bottomrule
\end{tabularx}
\end{table*}

Three of the four steps change their rating between columns, and the electromagnetic fault takes every one of the four ratings. The rating bins for high, medium, and low are 14, 6, and 5 points wide, while the state of knowledge moves elapsed time and expertise by up to 16 points in Table~\ref{tab:sens} and the profile moves the window of opportunity by up to 10. For major impact, the risk value of the same path then ranges from 1 to 4.

\subsection{Consequence for the TARA}
Under the attack-vector-based and CVSS-based approaches the resistance of the implementation does not enter the rating at all. Under the attack-potential-based approach it enters only through factor levels that the TARA team has to assume: how long an attack takes and what equipment it needs are properties of the implementation that cannot be read from the architecture. A feasibility rating of a hardware step is therefore a claim about implementation resistance. Rating the step as very low claims strong resistance, whether or not anyone has checked it. Two rules are missing: when such a claim may rest on argument, on documented evidence, or on test results, and how the step is rated while the evidence is not available. The method of Section~\ref{sec:method} provides both.

\section{Hardware-Aware TARA Methodology}
\label{sec:method}
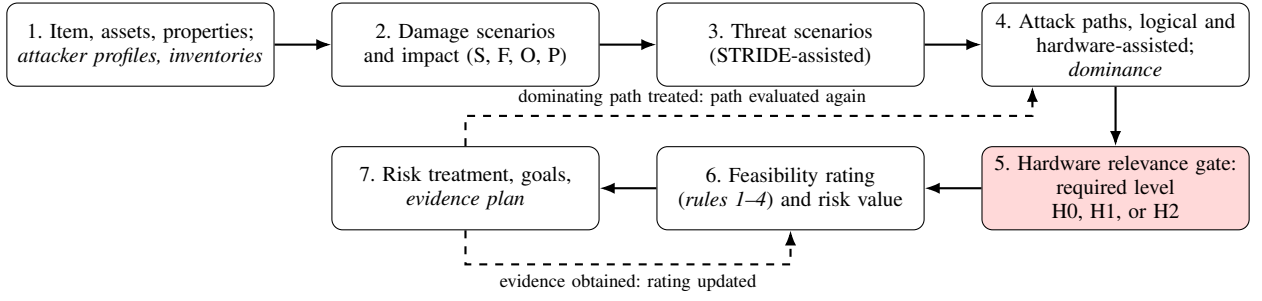
\begin{figure*}[t]
\centering
\begin{tikzpicture}[
    box/.style={draw,rounded corners,align=center,text width=3.3cm,minimum height=11mm,font=\footnotesize},
    new/.style={box,fill=red!15},
    arr/.style={-{Latex[length=2mm]},thick}
]
\node[box] (asset) at (0,0) {1. Item, assets, properties;\\\emph{attacker profiles, inventories}};
\node[box] (damage) at (4.3,0) {2. Damage scenarios\\and impact (S, F, O, P)};
\node[box] (threat) at (8.6,0) {3. Threat scenarios\\(STRIDE-assisted)};
\node[box] (paths) at (12.9,0) {4. Attack paths, logical and\\hardware-assisted; \emph{dominance}};
\node[new] (gate) at (12.9,-1.9) {5. Hardware relevance gate:\\required level H0, H1, or H2};
\node[box] (feas) at (8.6,-1.9) {6. Feasibility rating\\(\emph{rules 1--4}) and risk value};
\node[box] (treat) at (4.3,-1.9) {7. Risk treatment, goals,\\\emph{evidence plan}};
\draw[arr] (asset)--(damage);
\draw[arr] (damage)--(threat);
\draw[arr] (threat)--(paths);
\draw[arr] (paths)--(gate);
\draw[arr] (gate)--(feas);
\draw[arr] (feas)--(treat);
\draw[arr,dashed] (treat.south) -- ++(0,-0.45) -| node[pos=0.25,below,font=\scriptsize]{evidence obtained: rating updated} (feas.south);
\draw[arr,dashed] (treat.north) -- ++(0,0.4) -| node[pos=0.2,above,font=\scriptsize]{dominating path treated: path evaluated again} ([xshift=-11mm]paths.south);
\end{tikzpicture}
\caption{Hardware-aware TARA workflow. Steps~1 to~4, 6, and~7 are the activities of an ISO/SAE 21434 TARA; additions are set in italics, and Step~5 is new. Dashed arrows show the two iterations: a path is evaluated again after a path that dominates it has been treated, and a no-credit rating is replaced when evidence becomes available.}
\label{fig:workflow}
\end{figure*}

The method extends an existing ISO/SAE 21434-compatible TARA and leaves its impact scales, risk matrix, and treatment options unchanged. Fig.~\ref{fig:workflow} shows the workflow. The extension adds attacker profiles and two inventories to Step~1, hardware-assisted paths and a dominance check to Step~4, the hardware relevance gate as Step~5, and rating rules for hardware steps to Step~6. The gate is applied to each hardware-assisted attack path of a threat scenario. Its results are then collected per component, because evidence is requested and tests are planned per component.

\subsection{Step 1: Assets, Properties, and Attacker Profiles}
Hardware attacks are not introduced as generic threats to every microcontroller. The analysis starts from assets whose protection depends on an implementation that could leak, be faulted, or be reached through a debug or test interface:
\begin{itemize}
    \item cryptographic keys for in-vehicle authentication, diagnostics, secure boot, update verification, or vehicle-to-everything communication;
    \item boot state, rollback counters, and root-of-trust decisions;
    \item privilege state, debug authorization, and lifecycle configuration;
    \item decisions of safety and security monitors and integrity-critical calibration data;
    \item proprietary algorithms or models whose confidentiality has financial value.
\end{itemize}
For each asset the required cybersecurity property is recorded, for example ``the integrity of the boot-verification decision'' instead of ``the ECU is vulnerable to fault injection.'' Two inventories are added because Steps~4 and~5 depend on them. The \emph{key inventory} states for each secret whether it is unique per unit or shared across units and where it is stored and used. The \emph{decision inventory} states for each boot, update, debug, or diagnostic authorization decision where it is taken and which controls depend on it. The attacker profiles in scope (Table~\ref{tab:profiles}) are declared together with the item definition.

\subsection{Step 2: Damage Scenarios and Impact}
Damage scenarios are defined at the vehicle or stakeholder level and are independent of the attack technique. Disclosure of a key has a safety impact only if the key enables an attack path to a safety-relevant function, and a sophisticated laboratory technique against a low-value asset remains a low-impact scenario. Impact is rated from the consequence and never from the perceived severity of the technique.

\subsection{Step 3: Threat Scenarios}
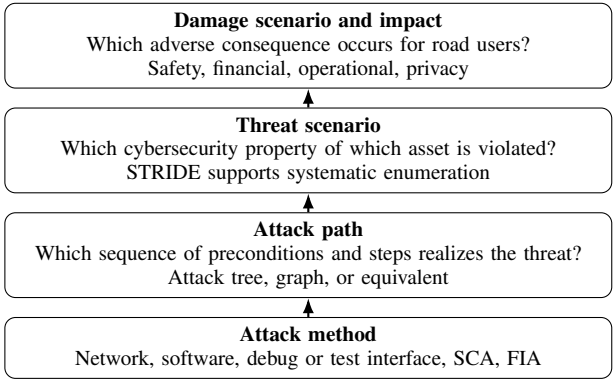
\begin{figure}[t]
\centering
\begin{tikzpicture}[
    node distance=2.5mm,
    box/.style={draw,rounded corners,align=center,text width=0.88\columnwidth,minimum height=7mm,font=\footnotesize},
    arr/.style={-{Latex[length=2mm]},thick}
]
\node[box] (damage) {\textbf{Damage scenario and impact}\\Which adverse consequence occurs for road users?\\Safety, financial, operational, privacy};
\node[box,below=of damage] (threat) {\textbf{Threat scenario}\\Which cybersecurity property of which asset is violated?\\STRIDE supports systematic enumeration};
\node[box,below=of threat] (path) {\textbf{Attack path}\\Which sequence of preconditions and steps realizes the threat?\\Attack tree, graph, or equivalent};
\node[box,below=of path] (method) {\textbf{Attack method}\\Network, software, debug or test interface, SCA, FIA};
\draw[arr] (method) -- (path);
\draw[arr] (path) -- (threat);
\draw[arr] (threat) -- (damage);
\end{tikzpicture}
\caption{Abstraction levels used in the method. SCA, FIA, and debug-interface abuse are attack methods within an attack path, STRIDE structures threat scenarios, and the impact categories apply to damage scenarios.}
\label{fig:levels}
\end{figure}

SCA, FIA, and debug-interface abuse are attack methods. They belong to the attack path and not to the threat scenario (Fig.~\ref{fig:levels}). A threat scenario states the violated property of an asset, for example ``confidentiality of the diagnostic authentication key is lost,'' and not ``power analysis is performed.''

STRIDE~\cite{shostack2014threat,abuabed2023stride} serves as a coverage checklist for threat scenarios. Table~\ref{tab:stride} shows which categories the three attack classes typically realize. SCA realizes information disclosure; spoofing or elevation of privilege follow only when the recovered secret is used. FIA realizes tampering, denial of service, and elevation of privilege directly. It also realizes information disclosure directly: differential fault analysis recovers a key from faulty cipher outputs~\cite{hou2024cryptography,baksi2023survey}, the same principle has been applied to the parameters of neural-network models~\cite{breier2022sniff}, and a fault on a read-out or access check exposes protected memory~\cite{milburn2018glitches,wouters2020dst80}. Debug access gives read, write, and execution control and therefore realizes most categories directly. STRIDE classifies the violated property and says nothing about impact: disclosure of a diagnostic key may have no privacy impact and a major operational or safety impact. The impact always comes from the damage scenario of Step~2.

\begin{table}[t]
\centering
\caption{Typical relationship between hardware attack classes and STRIDE categories. \yes: direct realization; \maybe: enabling or secondary step; \no: generally not a direct realization. Context can change individual entries.}
\label{tab:stride}
\scriptsize
\begin{tabular}{lccc}
\toprule
\textbf{STRIDE category} & \textbf{SCA} & \textbf{FIA} & \textbf{Debug} \\
\midrule
Spoofing & \maybe & \maybe & \maybe \\
Tampering & \no & \yes & \yes \\
Repudiation & \maybe & \maybe & \maybe \\
Information disclosure & \yes & \yes & \yes \\
Denial of service & \no & \yes & \yes \\
Elevation of privilege & \maybe & \yes & \yes \\
\bottomrule
\end{tabular}
\end{table}

\subsection{Step 4: Attack Paths and Path Dominance}
\label{sec:paths}
For each threat scenario $t$ the analyst identifies the set $\mathcal{P}(t)$ of attack paths that realize it. A path is an ordered sequence
\begin{equation}
P = \langle s_0, s_1, \ldots, s_n \rangle
\label{eq:path}
\end{equation}
from an entry condition $s_0$ to the asset violation $s_n$. A step that is realized by SCA, FIA, or a debug or test interface is a \emph{hardware step}, and a path with at least one hardware step is \emph{hardware-assisted}. Typical forms are:
\begin{itemize}
    \item access to a unit $\rightarrow$ repeated cryptographic operation $\rightarrow$ SCA $\rightarrow$ key disclosure $\rightarrow$ impersonation of an ECU;
    \item access to a unit $\rightarrow$ reachable debug or boot-mode interface $\rightarrow$ read or write access $\rightarrow$ disclosure or modification of firmware and keys;
    \item remote software compromise $\rightarrow$ access to an on-chip measurement or regulation resource $\rightarrow$ leakage or fault $\rightarrow$ violation of a security property.
\end{itemize}
A hardware-assisted path is added for every asset in the inventories of Step~1 that a hardware step can reach under at least one declared profile. Attack trees or attack graphs~\cite{lautenbach2021heavens,zelle2022threatsurf} can hold these paths without change.

Paths to the same consequence compete. An analyst who moves directly from ``valuable key'' to ``SCA testing'' may overlook that the key can be read through an open debug port. We define this competition on the attack-potential factors. For a path $P$, let
\begin{equation}
\vP(P) = (v_{\mathrm{ET}}, v_{\mathrm{SE}}, v_{\mathrm{KN}}, v_{\mathrm{WO}}, v_{\mathrm{EQ}})
\label{eq:vector}
\end{equation}
be the levels of elapsed time, specialist expertise, knowledge of the item, window of opportunity, and equipment required for the complete path, and let $C(P)$ be the set of damage scenarios that $P$ can cause.

\emph{Definition 1 (dominance):} A path $P'$ dominates a path $P$, written $P' \preceq P$, if $C(P') \supseteq C(P)$ and $\vP(P') \leq \vP(P)$ in every component.

The comparison is ordinal. It does not depend on the numeric values of Table~\ref{tab:apvalues} and can be made before any path is rated. Every feasibility rating $\varphi$ that cannot increase when a factor level rises gives $\varphi(P') \geq \varphi(P)$. With the aggregation that the standard gives as an example,
\begin{equation}
\varphi(t) = \max_{P \in \mathcal{P}(t)} \varphi(P),
\label{eq:max}
\end{equation}
a dominated path cannot raise the risk of any damage scenario it reaches as long as the dominating path exists.

\emph{Definition 2 (remediable and inherent dominance):} The dominance of $P'$ over $P$ is \emph{remediable} if a cybersecurity control in the scope of the item removes $P'$ or raises $\vP(P')$ until $P'$ no longer dominates $P$. Otherwise it is \emph{inherent}.

An open debug port that exposes the same key as an SCA path dominates that path remediably: the port has to be closed first, and the SCA path may then become the easiest remaining one. A fault attack that only forces an ECU into resets is dominated inherently by flooding the CAN segment of that ECU, which the same attacker can reach: priority-based arbitration lets any node on the segment keep the others from transmitting~\cite{koscher2010experimental}, the loss of availability is the same, and the property belongs to the bus protocol and not to a missing control of the item. Dominance is also treated as inherent when the risk of $P'$ has been retained, because $P$ cannot then carry a higher risk and is covered by the same claim. Remediable dominance orders the work. Inherent dominance is the precise form of the argument that a physical attack adds nothing to what the attacker can do anyway, and it holds only if the consequences are the same.

\subsection{Step 5: Hardware Relevance Gate}
\label{sec:gate}
The gate decides how the resistance of a hardware step has to be supported before the feasibility rating may rely on it. It uses the impact from Step~2 and structural properties of the path. It does not use the feasibility rating, which is the quantity in question, and it does not produce a second risk value.

\subsubsection{Baseline}
Three facts are established for every ECU in scope, independent of the level, because the gate inputs depend on them and because establishing them costs little compared to a specialist hardware evaluation: the key inventory, the decision inventory, and the production state of all debug, test, programming, and boot-mode interfaces, verified for each production-relevant ECU variant and lifecycle configuration (Table~\ref{tab:evidence}, first column).

\subsubsection{Inputs}
For each hardware-assisted path $P$ of a threat scenario $t$, five inputs are determined.
\begin{itemize}
    \item \emph{Exposure} $x(P) \in \{0,1\}$. The path is exposed, $x=1$, if for at least one profile in scope the attacker can obtain \emph{access} to the component or to another unit of the same type, can \emph{trigger} the sensitive operation or decision often enough or at a useful time, and can \emph{observe} the leakage, use the effect of the fault, or reach and activate the interface.
    \item \emph{Dominance} $d(P) \in \{\text{none}, \text{remediable}, \text{inherent}\}$ according to Definitions~1 and~2.
    \item \emph{Impact} $I(t)$: the highest impact rating among the damage scenarios linked to $t$.
    \item \emph{Leverage} $\ell(P) \in \{\text{local}, \text{extended}\}$. Leverage is extended if the result is \emph{scalable} (S): the violated asset is shared across units, such as a class-wide key or a secret embedded in firmware, or the result enables attacks on other units without physical access. It is also extended if the step defeats a \emph{trust anchor} (T): a decision on which other controls of the unit depend, such as secure boot, update authentication, or debug authorization. Otherwise the effect is confined to the attacked unit and leverage is local.
    \item \emph{Known attack} $k(P) \in \{0,1\}$: an attack on the same component family or security primitive in a comparable configuration has been published. A known attack supports $x=1$ only if the demonstrated access and trigger conditions are compatible with at least one profile in scope and with the configuration under analysis.
\end{itemize}
Exposure, dominance, and leverage are answered from the architecture, the inventories, and the declared profiles. Together with the available evidence, the inputs make explicit the questions that an analyst would ask informally: how valuable the asset is, whether the attacker can get access, whether the operation can be triggered and its effect observed, what the step unlocks, and what is known about resistance.

\begin{figure}[t]
\centering
\begin{tikzpicture}[
    font=\footnotesize,
    proc/.style={draw,rounded corners,align=center,text width=4.6cm,minimum height=7mm,inner sep=2pt},
    dec/.style={draw,diamond,aspect=2.4,align=center,inner sep=0.5pt,text width=1.75cm},
    res/.style={draw,rounded corners,fill=gray!15,align=center,text width=1.8cm,minimum height=6mm,inner sep=2pt},
    arr/.style={-{Latex[length=1.8mm]},thick}
]
\node[proc] (start) at (0,0) {Hardware-assisted path $P$\\of threat scenario $t$};
\node[dec] (exp) at (0,-1.3) {Exposed?\\$x(P)=1$};
\node[dec] (dom) at (0,-2.95) {Dominance\\$d(P)$};
\node[proc] (lvl) at (0,-4.55) {Level from $I(t)$ and $\ell(P)$ (Table~\ref{tab:matrix});\\H1 becomes H2 if $k(P)=1$};
\node[dec] (ev) at (0,-6.05) {Required\\evidence?};
\node[res] (h0a) at (3.3,-1.3) {H0:\\record claim};
\node[res] (h0b) at (3.3,-2.95) {Inherent: H0,\\record claim};
\node[res] (treat) at (-3.3,-2.95) {Remediable:\\treat the dominating path first};
\node[res] (rated) at (3.3,-6.05) {Available: rate with evidence (rule~4)};
\node[proc] (default) at (0,-7.55) {No-credit rating (rule~3); request evidence (H1)\\or plan evaluation (H2)};
\draw[arr] (start) -- (exp);
\draw[arr] (exp) -- node[left]{yes} (dom);
\draw[arr] (exp) -- node[above]{no} (h0a);
\draw[arr] (dom) -- node[left]{none} (lvl);
\draw[arr] (dom) -- (h0b);
\draw[arr] (dom) -- (treat);
\draw[arr] (treat.north) |- (exp.west);
\draw[arr] (lvl) -- node[left]{H1 or H2} (ev);
\draw[arr] (lvl.east) -| node[pos=0.25,above]{H0} (h0b.south);
\draw[arr] (ev) -- (rated);
\draw[arr] (ev) -- node[left]{missing} (default);
\end{tikzpicture}
\caption{Decision rule of the hardware relevance gate for one hardware-assisted attack path.}
\label{fig:gate}
\end{figure}
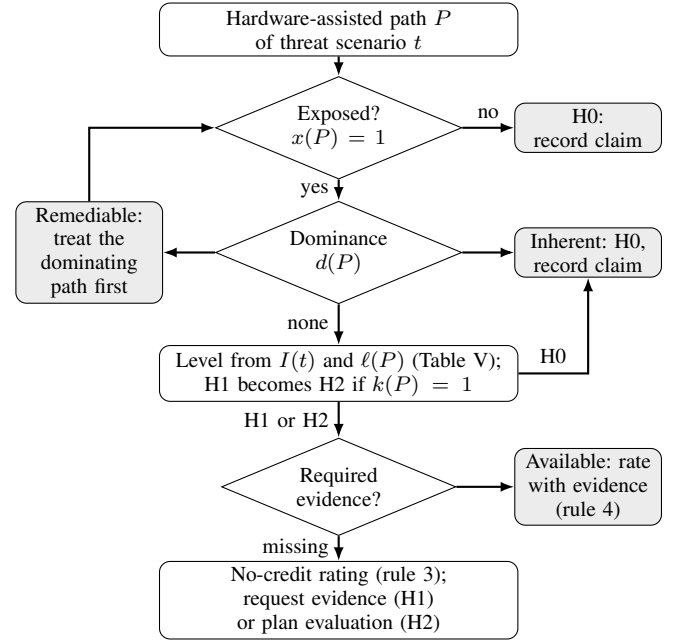

\subsubsection{Decision rule}
Fig.~\ref{fig:gate} shows the rule. A path that is not exposed or that is inherently dominated is H0. A remediably dominated path is deferred: the dominating path is treated first, and the path is then evaluated again. Otherwise Table~\ref{tab:matrix} gives the level from impact and leverage, and a known attack raises H1 to H2.

\begin{table}[t]
\centering
\caption{Default mapping of impact and leverage to the required level for an exposed path that is not dominated. A known attack ($k=1$) raises H1 to H2.}
\label{tab:matrix}
\scriptsize
\begin{tabular}{lcc}
\toprule
 & \multicolumn{2}{c}{\textbf{Leverage} $\ell(P)$} \\
\cmidrule(lr){2-3}
\textbf{Impact} $I(t)$ & local & extended (S or T) \\
\midrule
Negligible & H0 & H0 \\
Moderate & H0 & H1 \\
Major & H1 & H2 \\
Severe & H1 & H2 \\
\bottomrule
\end{tabular}
\end{table}

The table is a default that an organization may calibrate, as with the CAL example of Annex~E. Its logic is that the attack vector, which Annex~E uses as the second determinant, carries no information for hardware steps, whereas leverage separates an attack on one unit from an attack on a fleet or on the root of trust. Severe impact with local leverage is H1 by default, because the attacker must have had physical access to the vehicle that is affected; an organization that does not accept this argument can raise the cell. Three properties should survive any calibration: the level never decreases when impact or leverage increases, negligible impact never requires evidence, and demonstrated evidence is required only where the result extends beyond the attacked unit or defeats a trust anchor, or where a comparable attack is already published.

\subsubsection{Levels and evidence}
Table~\ref{tab:gate} defines the levels. H0 does not state that the hardware is secure. It records, as a cybersecurity claim, why no dedicated evidence is needed, together with the assumption that would reopen the question. H1 and H2 do not state that an attack will succeed. They state that an assumed resistance is not sufficient. Evidence is \emph{documented} (E1) if the resistance of the specific primitive or decision, in the configuration used, is supported by the result of an evaluation that the supplier or a third party provides in summary form: a leakage-assessment or fault-injection evaluation summary, or a certification report whose scope covers that primitive and configuration. It is \emph{demonstrated} (E2) if it is supported by test results for production-representative hardware, firmware, and lifecycle state that the TARA team can review, whoever produced them. A design description of countermeasures is neither. It states that protection exists and is needed to judge whether an evaluation applies, but it does not show how strong the protection is and earns no credit under rule~4. Published evaluations show why: single-bit parity protection of a cipher was ineffective against laser fault injection, which typically flips several adjacent bits, and non-linear codes with better theoretical detection bounds brought little improvement over simpler linear ones~\cite{breier2017attacks}. A shuffling countermeasure against SCA was found to leak through the generation of its own permutation; the redesign in~\cite{puvskavc2025make} removes this leak and was evaluated by correlation power analysis on an ARM Cortex-M4 microcontroller. Table~\ref{tab:evidence} lists typical evidence per attack class; leakage assessment for SCA can follow ISO/IEC 17825 or the test vector leakage assessment methodology~\cite{iso17825,goodwill2011tvla}. Verifying the production debug state is part of the baseline, whereas testing whether the lock can be defeated is E2 evidence.

\begin{table*}[t]
\centering
\caption{Levels of the hardware relevance gate. A level states what must support the feasibility rating of a hardware step. It is not a risk value.}
\label{tab:gate}
\scriptsize
\begin{tabularx}{\textwidth}{p{0.6cm}p{2.5cm}Y p{3.3cm} Y}
\toprule
\textbf{Level} & \textbf{Resistance of the hardware step} & \textbf{Assigned when} & \textbf{Evidence accepted} & \textbf{Action while the evidence is missing} \\
\midrule
H0 & may be argued & Path not exposed; path inherently dominated; negligible impact; moderate impact with local leverage. & Recorded rationale with its assumptions (cybersecurity claim). & None beyond the baseline. The decision is reopened by the triggers of Step~7. \\
\addlinespace
H1 & must be documented & Moderate impact with extended leverage; major or severe impact with local leverage. & E1 or E2. & Request the evidence through the cybersecurity interface agreement. Use the no-credit rating (rule~3). \\
\addlinespace
H2 & must be demonstrated & Major or severe impact with extended leverage; any H1 path for which an attack on the component family is published. & E2 only. & Plan an evaluation on a production-representative unit and derive resistance requirements. Use the no-credit rating (rule~3). \\
\bottomrule
\end{tabularx}
\end{table*}

\begin{table*}[t]
\centering
\caption{Baseline facts and typical evidence per attack class.}
\label{tab:evidence}
\scriptsize
\begin{tabularx}{\textwidth}{p{0.75cm}YYY}
\toprule
\textbf{Class} & \textbf{Baseline (every ECU in scope)} & \textbf{E1: documented} & \textbf{E2: demonstrated} \\
\midrule
SCA & Key inventory: which secrets are shared across units, and where each secret is stored and used. & Leakage-assessment summary or certification report that covers the primitive processing the asset and its configuration, supported by the description of the countermeasures and of the limits on key use. & Leakage assessment and attack-based evaluation on production-representative hardware and firmware, with a report that can be reviewed. \\
\addlinespace
FIA & Decision inventory: boot, update, debug, and diagnostic authorization decisions, and which controls depend on each. & Fault-injection evaluation summary or certification report that covers the decision and its configuration, supported by the design description of redundant and fail-secure checks, fault detectors, and the reaction to detected faults. & Fault-injection campaign against the specific decision (voltage or clock, electromagnetic; optical if the attacker model requires it), reporting success rate and repeatability. \\
\addlinespace
Debug & Production state of debug, test, programming, and boot-mode interfaces, verified for each production-relevant variant and lifecycle configuration. & Evaluation summary for the lock and for the unlock or authentication mechanism in the production lifecycle state, supported by the lifecycle-state design, its key management, and the service procedures that re-enable access. & Adversarial test of the lock: unlock attempts, lifecycle regression, residual interfaces, and fault injection against the lock. \\
\bottomrule
\end{tabularx}
\end{table*}

\subsection{Step 6: Feasibility Rating of Hardware Steps}
\label{sec:rating}
The path is rated with the method of the organization. If that method is attack-vector-based or CVSS-based, the result for a physical path is fixed (Section~\ref{sec:analysis}), so hardware-assisted paths at H1 or H2 should be rated with an attack-potential-based method. Four rules apply.

\emph{Rule 1 (follower):} The step is rated for an attacker who reproduces a known attack, unless evidence shows that the characterization has to be repeated for each unit. Vehicles can be in service for well over a decade, and procedures, parameters, and extracted class-wide secrets become available to followers within that time. This corresponds to the exploitation phase in~\cite{jil2024attackpotential}.

\emph{Rule 2 (profile):} The window of opportunity is that of the profile with the widest access among those that can cause the damage scenario. For scalable assets this is the donor-unit profile with an unlimited window.

\emph{Rule 3 (no credit):} While evidence at the required level is unavailable, no credit is given for implementation-specific resistance. For paths without a known attack, the follower reference values of Table~\ref{tab:sens} are used as a conservative no-credit default; when a corresponding attack is already known, Rule 1 applies.

\emph{Rule 4 (evidence):} When the required evidence is available, the factors are set to the levels that the evidence supports, and the rating refers to the evidence.

Obscurity is not rated as resistance: knowledge of the item is rated as public unless evidence shows that the attacker needs information that is controlled and cannot be recovered from a donor unit. The repeatability of a fault and the number of traces needed for SCA enter through elapsed time. Scalability enters through leverage and, under rule~2, through the window of opportunity, and is not added to the impact a second time.

\subsection{Step 7: Risk Treatment and Evidence Plan}
Treatment is expressed as cybersecurity goals and requirements at the appropriate architectural level, and hardware countermeasures are only one option. Architectural measures often change the gate inputs: key diversification per vehicle turns extended into local leverage, asymmetric authentication removes a shared secret from the ECU, and compartmentalization lowers the impact. For SCA, further measures are leakage-resistant implementations, hardened security modules, short-lived credentials, and limits on key use. For FIA, they are redundant and fail-secure checks, fault detectors, and safe-state handling. For debug interfaces, they are permanent disabling where field access is not needed, authenticated debug where it is, explicit lifecycle states, and protection against re-enabling.

The level determines the evidence plan. H1 evidence is requested from the supplier through the cybersecurity interface agreement, and H2 evaluations are planned as verification and validation activities. An H0 rationale is recorded as a cybersecurity claim. Three events reopen a decision: a change of the architecture or of the attacker profiles in scope, a new entry in the key or decision inventory, and the publication of an attack on a component family used in the item, which sets $k=1$. The last event is detected by cybersecurity monitoring throughout the support period.

\subsection{Relation to CAL and TAF}
\label{sec:cal}
The H-levels grade assurance, as CALs do, and the two should not be confused. A CAL applies to a component or a cybersecurity goal and scales the rigor of all cybersecurity activities; the Annex~E example derives it from impact and attack vector and therefore assigns the lowest levels to physical paths (Section~\ref{sec:analysis}). An H-level applies to one hardware step of one attack path and fixes one thing only: the kind of evidence that must support its feasibility rating. It is a refinement for a class of attack steps for which the vector is uninformative, not a competing scale. In an organization that uses CALs, the gate can tailor verification: an H2 path identifies a place where testing with hardware techniques is required even though the CAL obtained from the attack vector is low. TAF fits the same picture. A TAF for a hardware-assisted path, for example ``low or below,'' is a resistance target. The H-level states whether reaching it may be documented or must be demonstrated, and rule~3 states how the path is rated until then.

\section{Application to a Reference Architecture}
\label{sec:example}
\subsection{Item and Assumptions}
\begin{figure}[t]
\centering
\begin{tikzpicture}[
    font=\sffamily\scriptsize,
    ecu/.style={draw,rounded corners,align=center,minimum height=8mm,text width=1.45cm,fill=white},
    gw/.style={draw,thick,rounded corners,align=center,minimum height=17mm,text width=2.5cm,fill=gray!12},
    bus/.style={thick},
    att/.style={-{Latex[length=1.6mm]},red,thick}
]
\node[gw] (gw) at (0,0) {\textbf{Gateway (GW)}\\microcontroller with HSM\\$K_D$ in firmware, $K_S$ in HSM\\secure boot, debug port};
\node[ecu] (tcu) at (-3.05,1.05) {TCU\\(cellular)};
\node[ecu] (obd) at (-3.05,-1.05) {OBD\\connector};
\node[ecu] (pcm) at (3.05,1.05) {PCM\\(speed limiter)};
\node[ecu] (chs) at (3.05,0) {Chassis ECUs};
\node[ecu] (ilm) at (3.05,-1.05) {ILM\\(interior light)};
\draw[bus] (tcu.east) -- ([yshift=5mm]gw.west);
\draw[bus] (obd.east) -- ([yshift=-5mm]gw.west);
\draw[bus] ([yshift=5mm]gw.east) -- (pcm.west);
\draw[bus] (gw.east) -- (chs.west);
\draw[bus] ([yshift=-5mm]gw.east) -- (ilm.west);
\node[font=\sffamily\tiny,red,align=center] (p1) at (-0.95,-1.75) {P1a--c, P2, P3};
\draw[att] (p1.north) -- ([xshift=-9.5mm]gw.south);
\node[font=\sffamily\tiny,red] (p5) at (-3.05,1.95) {P5};
\draw[att] (p5.south) -- (tcu.north);
\node[font=\sffamily\tiny,red] (p6) at (3.05,1.95) {P6};
\draw[att] (p6.south) -- (pcm.north);
\node[font=\sffamily\tiny,red] (p4) at (3.05,-1.95) {P4};
\draw[att] (p4.north) -- (ilm.south);
\end{tikzpicture}
\caption{Reference item. Arrows mark where the attack paths of Table~\ref{tab:example} apply their hardware step; P5 starts from compromised software on the TCU.}
\label{fig:item}
\end{figure}
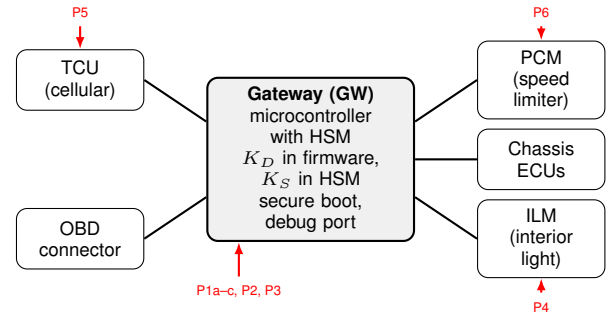

The item (Fig.~\ref{fig:item}) is a gateway ECU (GW) with a microcontroller that contains a hardware security module (HSM). The GW routes between a telematics unit (TCU), the chassis and powertrain domain, and the body domain. It authenticates safety-relevant frames with SecOC, verifies its firmware at boot, and offers diagnostics and reprogramming through the on-board diagnostics (OBD) connector. A powertrain control module (PCM) holds a speed-limiter calibration, and an interior-light module (ILM), a LIN slave whose commands are not authenticated, serves as a low-impact comparison. The architecture is representative and does not describe a specific product. The initial design has the following properties.
\begin{enumerate}
    \item[(a)] Privileged diagnostic sessions are authorized by a challenge-response protocol with a symmetric secret $K_D$ that is identical for all vehicles of the type and is embedded in the GW firmware. The encryption runs in software.
    \item[(b)] The SecOC keys $K_S$ are unique per vehicle and are stored and used only inside the HSM. The supplier provides leakage-assessment evidence that covers the AES-CMAC implementation of this HSM in the configuration used.
    \item[(c)] Boot verification uses the HSM. The microcontroller supplier makes no statement about fault resistance, while fault attacks on the bootloaders of comparable gateway microcontrollers are published~\cite{pozzobon2023fuzzy}.
    \item[(d)] The production state of the GW debug port has not been verified.
    \item[(e)] The GW microcontroller has no clock or voltage regulator under software control and no analog input that shares a supply with the HSM.
    \item[(f)] The PCM protects its boot-mode loader with a password check, for which a fault attack on a comparable microcontroller family is published~\cite{oflynn2020bam}.
\end{enumerate}
All five profiles of Table~\ref{tab:profiles} are in scope. The impact ratings are assumptions of the kind an organization could plausibly make, and the example risk matrix of Annex~H is used.

\subsection{Results}
Table~\ref{tab:example} lists five threat scenarios and eight hardware-assisted paths with their gate decisions, and Table~\ref{tab:ratings} gives the feasibility ratings.

\begin{table*}[t]
\centering
\caption{Gate decisions for the reference item (Steps~4, 5, and~7). All paths except P5 are exposed. Leverage: local, or extended by scalability (S) or by a trust anchor (T). Evidence: E1, documented.}
\label{tab:example}
\scriptsize
\setlength{\tabcolsep}{4pt}
\begin{tabularx}{\textwidth}{p{0.5cm} Y p{1.0cm} p{1.9cm} p{1.0cm} p{0.8cm} p{0.6cm} p{0.95cm} p{3.9cm}}
\toprule
\textbf{Path} & \textbf{Threat scenario and hardware step} & \textbf{Impact} & \textbf{Exposure, dominance} & \textbf{Leverage} & \textbf{Known attack} & \textbf{Level} & \textbf{Evidence} & \textbf{Outcome} \\
\midrule
P1a & T1: confidentiality of the class-wide diagnostic secret $K_D$. Read-out through the debug port of a donor unit. & Major & Not dominated & Ext. (S) & No & H2 & None & Port locked and lock verified (baseline); class-wide secret removed. \\
\addlinespace
P1b & T1. Voltage glitch on read-out protection or on the diagnostic memory-read check. & Major & Dominated by P1a (remediable) & Ext. (S) & No & H2 & None & Deferred until P1a is closed, then removed with the class-wide secret. \\
\addlinespace
P1c & T1. Power analysis of the software cipher. & Major & Dominated by P1a (remediable) & Ext. (S) & No & H2 & None & As P1b. \\
\addlinespace
P2 & T2: confidentiality of the per-vehicle SecOC keys $K_S$. Power analysis of the HSM. & Severe & Not dominated & Local & No & H1 & E1 & Evaluation summary of the supplier reviewed for applicability; claim recorded. \\
\addlinespace
P3 & T3: integrity of the GW boot decision. Electromagnetic or voltage fault during signature verification. & Severe & Not dominated & Ext. (T) & Yes & H2 & None & Fault-injection evaluation; redundant, fail-secure verification. \\
\addlinespace
P4 & T4: integrity and availability of the interior-light function. Any hardware step on the ILM. & Moderate & Inherently dominated & Local & No & H0 & -- & Rationale recorded as claim. \\
\addlinespace
P5 & T2 through software-mediated SCA or FIA from the TCU. & Severe & Not exposed & -- & -- & H0 & -- & Assumption (e) recorded as claim. \\
\addlinespace
P6 & T5: integrity of the speed-limiter calibration, attacked by the owner. Electromagnetic fault on the boot-mode password check of the PCM. & Major & Not dominated & Ext. (T) & Yes & H2 & None & Fault-injection evaluation; fail-secure password check. \\
\bottomrule
\end{tabularx}
\end{table*}

\begin{table}[t]
\centering
\caption{Feasibility ratings and risk values of the physical paths of Table~\ref{tab:example} (Step~6). $\vP$: levels (ET, SE, KN, WO, EQ); $\Sigma$: sum; $R$: risk value; $R_{\mathrm{AV}}$: risk value under the attack-vector-based approach, which the CVSS-based approach reproduces.}
\label{tab:ratings}
\scriptsize
\setlength{\tabcolsep}{3.5pt}
\begin{tabularx}{\columnwidth}{p{0.55cm} Y p{1.55cm} c p{0.95cm} c c}
\toprule
\textbf{Path} & \textbf{Basis and profile} & $\vP(P)$ & $\Sigma$ & \textbf{Rating} & $R$ & $R_{\mathrm{AV}}$ \\
\midrule
P1a & Port found open; A4 & (0, 3, 0, 0, 0) & 3 & High & 4 & 1 \\
P1b & No credit; A4 & (1, 3, 0, 0, 4) & 8 & High & 4 & 1 \\
P1c & No credit; A4 & (1, 3, 0, 0, 4) & 8 & High & 4 & 1 \\
P2 & E1 evidence; A1 & (17, 6, 0, 0, 4) & 27 & Very low & 2 & 2 \\
P2 & No credit, for comparison; A1 & (1, 3, 0, 0, 4) & 8 & High & 5 & 2 \\
P3 & No credit; A1 & (1, 6, 0, 0, 4) & 11 & High & 5 & 2 \\
P4 & Not rated (H0) & -- & -- & -- & -- & 1 \\
P6 & No credit; A1 & (1, 6, 0, 0, 4) & 11 & High & 4 & 1 \\
\bottomrule
\end{tabularx}
\end{table}

\emph{T1, diagnostic secret:} The key inventory marks $K_D$ as shared across units, so leverage is scalable and, with major impact, all three paths are H2. The baseline check finds the debug port open (P1a). P1a dominates P1b and P1c remediably, since it needs no more time, expertise, or equipment and reaches the same damage scenario. It is treated first, by locking the port and verifying the lock on a production sample. P1b and P1c are then evaluated again. No evidence exists for either, so rule~3 gives no credit and rule~2 applies the unlimited window of the donor-unit profile. The no-credit rating is high feasibility with a risk value of 4. Demonstrating the resistance of a software encryption and of the read-out protection would require two evaluations. The treatment chosen instead removes the asset: with certificate-based authentication only a public verification key remains in the ECU, which ends T1. The verification decision becomes a new entry in the decision inventory and is handled like T3.

\emph{T2, SecOC keys:} The keys are unique per vehicle and no trust anchor is defeated, so leverage is local and the level is H1 despite the severe impact. The leakage-assessment summary of the supplier is E1 evidence, because it covers the HSM implementation and the CMAC configuration that SecOC uses. After this applicability has been reviewed, the path is rated with the factor levels that the summary supports, in this example an elapsed time of up to six months and expert-level skill, which gives very low feasibility and a risk value of 2. A statement that the HSM contains countermeasures against power analysis would not have been sufficient. The no-credit rating would then have applied, with high feasibility and a risk value of 5. If an SCA attack on this HSM family were published, $k=1$ would raise the level to H2 and the summary would no longer suffice.

\emph{T3, boot decision:} Secure boot is a trust anchor, the impact is severe, and a comparable attack is known, so the level is H2. No evidence of fault resistance exists. The no-credit rating for the owner profile is high feasibility with a risk value of 5. The treatment combines redundant, fail-secure verification with a fault-injection evaluation of the boot decision on a production-representative GW. The result of the evaluation replaces the no-credit rating.

\emph{T4, interior light:} Impact is moderate and leverage is local, which gives H0. The path is also inherently dominated. An attacker who reaches the module also reaches its LIN pin and can issue the same lighting commands or block the bus with standard equipment, and the risk of this unauthenticated bus access has been retained because of the moderate impact. The rationale is recorded with the assumption that the ILM holds no shared secret and has no authority beyond the lighting function.

\emph{T2 through software-mediated paths:} By assumption~(e) no measurement or perturbation resource can be reached by software, so the path is not exposed. It is H0, with~(e) recorded as the claim to be revisited if the microcontroller changes.

\emph{T5, owner tampering:} The owner profile gives an unlimited window. The boot-mode password is a trust anchor for the flash contents and a comparable attack is published, so the level is H2. The no-credit rating is high feasibility with a risk value of 4.

\subsection{Comparison With the Unextended TARA}
Under the attack-vector-based or the CVSS-based approach, all seven physical paths are rated very low and their risk values are 1 or 2 (column $R_{\mathrm{AV}}$ of Table~\ref{tab:ratings}). An organization that reduces risks from a value of 3 upward (a common threshold, in our experience) would have derived no requirement from any of them, including the open debug port in front of a fleet-wide secret. The Annex~E example would have assigned CAL1 to T1 and T5 and CAL2 to T2 and T3. Under the attack-potential-based approach without the rules of Step~6, the outcome depends on the analyst: the fault attack of P3, for instance, is very low when rated for a first attacker and medium to high when rated for a follower (Table~\ref{tab:sens}). With the extension, four paths lead to treatment or evaluation at a risk value of 4, one at a risk value of 5, one is closed by reviewed supplier evidence, and two are excluded with recorded assumptions. The additional effort before any testing consists of the two inventories and the verification of the debug state for each production-relevant ECU variant and lifecycle configuration.

\section{Retrospective Structural Analysis of Published Attacks}
\label{sec:retro}
\subsection{Procedure}
We examined ten published attacks on automotive components (Table~\ref{tab:retro}). In each of them, SCA, FIA, or debug access was demonstrated on an automotive component or on an implementation of an automotive protocol, with enough detail to identify the asset and the hardware step. The selection is purposive and not exhaustive. It covers keyless entry, immobilizers, in-vehicle communication, diagnostics, bootloaders, and vehicle computers between 2008 and 2025. Characterization studies without an asset, such as~\cite{wiersma2017safety}, are not included.

The item definitions, damage scenarios, and impact ratings of the affected products are not available. The analysis therefore does not reproduce a TARA and assigns no risk values. We use the studies to examine the structural inputs of the method: for which profile the conditions of exposure were met, whether the hardware step attacks a decision or a cryptographic computation, which leverage the result has, and whether competing implementation-level paths exist. The gate is considered as it would have been at design time, that is, with $k=0$, and its outcome is stated as a function of the impact.

\begin{table*}[t]
\centering
\caption{Structural inputs of the method for ten published attacks. Target of the hardware step: D, an access-control or verification decision or an accessible interface; C, a cryptographic computation. Leverage: S, scalable; T, trust anchor.}
\label{tab:retro}
\scriptsize
\begin{tabularx}{\textwidth}{p{1.25cm} p{3.9cm} p{3.7cm} c p{1.45cm} Y}
\toprule
\textbf{Study} & \textbf{Component and asset} & \textbf{Hardware step} & \textbf{Target} & \textbf{Leverage} & \textbf{Reported means and result} \\
\midrule
\cite{eisenbarth2008keeloq}, 2008 & Remote keyless entry (KeeLoq); device keys and manufacturer key & Differential power analysis of transmitter and receiver & C & S & Ten power traces to clone a transmitter; with the manufacturer key, cloning from two eavesdropped messages. \\
\addlinespace[1.5pt]
\cite{garcia2016lock}, 2016 & Remote keyless entry of one manufacturer group; global master keys in ECU firmware & Firmware read-out through the programming interface & D & S & Standard programming tools; afterwards one eavesdropped signal suffices to clone a remote control. \\
\addlinespace[1.5pt]
\cite{herrewegen2018bonnet}, 2018 & Diagnostic access control of several ECUs; firmware & Firmware read-out through JTAG and BDM & D & S & Debug access to production ECUs. \\
\addlinespace[1.5pt]
\cite{milburn2018glitches}, 2018 & Instrument cluster; firmware & Voltage glitch on the diagnostic memory-read authorization & D & S & About three days and $3 \times 10^5$ glitch attempts for the complete firmware image. \\
\addlinespace[1.5pt]
\cite{oflynn2018securecan}, 2018 & AES-based authenticated CAN on a general-purpose microcontroller; bus keys & Correlation power analysis; fault attacks & C & depends on key sharing & Low-cost capture equipment. \\
\addlinespace[1.5pt]
\cite{oflynn2020bam}, 2020 & Production ECU of 2019; boot-mode password & Electromagnetic fault on the password check & D & T & In situ on the ECU, under conditions similar to a garage bench. \\
\addlinespace[1.5pt]
\cite{wouters2020dst80}, 2020 & Immobilizer ECUs of several manufacturers; key diversification schemes & Voltage glitch on firmware read-out protection; profiled SCA of key loading & D, C & S & Keys derivable from public transponder data or with three bytes of entropy. \\
\addlinespace[1.5pt]
\cite{ebina2023secoc}, 2023 & SecOC-compliant AES-CMAC; MAC key & SCA of the MAC computation & C & depends on key sharing & Key recovery from the MAC computation. \\
\addlinespace[1.5pt]
\cite{pozzobon2023fuzzy}, 2023 & Gateway ECUs of two manufacturers; update signature verification & Electromagnetic fault during verification & D & T, S & No hardware modification; code execution, extraction of firmware and secrets, debug re-enabled. \\
\addlinespace[1.5pt]
\cite{kuhnapfel2025three}, 2025 & Car computer of a connected electric vehicle; three subsystems & Voltage glitches (two subsystems), electromagnetic fault (one) & D & T & Low-cost glitching and non-invasive electromagnetic fault; all three subsystems compromised. \\
\bottomrule
\end{tabularx}
\end{table*}

\subsection{Results}
\emph{Exposure:} In all ten studies the hardware step was carried out on units in the hands of the researchers, which corresponds to the owner or the donor-unit profile (A1 and A4). Access, trigger, and observation were available without access to a vehicle in use by a third party.

\emph{Target of the hardware step:} In seven of the ten studies the attacker obtains code or privileged execution through a debug, programming, boot, diagnostic, or update mechanism. Two use an accessible debug or programming interface~\cite{garcia2016lock,herrewegen2018bonnet}, and five defeat an authorization or verification decision by a fault~\cite{milburn2018glitches,oflynn2020bam,wouters2020dst80,pozzobon2023fuzzy,kuhnapfel2025three}. Keys are then read from the extracted firmware or are no longer needed. Three studies recover a key from a cryptographic computation by SCA~\cite{eisenbarth2008keeloq,oflynn2018securecan,ebina2023secoc}. The set is purposive, so these counts describe the ten studies and are not an estimate of how often each kind of step occurs. They are consistent with the order that the baseline and the dominance check impose: debug and lifecycle state first, decisions second, cryptographic implementations third.

\emph{Competing paths:} One study shows remediable dominance directly. In~\cite{wouters2020dst80} the keys follow from weaknesses of the key diversification that were found in firmware extracted by a fault on the read-out protection, and the authors show that a profiled SCA of key loading remains possible without them.

\emph{Leverage:} In eight studies the step yields a class-wide secret or firmware (S) or defeats a trust anchor (T); in the other two the leverage depends on key sharing across vehicles. In~\cite{eisenbarth2008keeloq} and~\cite{garcia2016lock} a single manufacturer or master key allows a remote control to be cloned from one or two eavesdropped messages.

\emph{Gate outcome as a function of impact:} All ten paths need physical access for the hardware step, so the attack-vector-based and the CVSS-based approach rate them very low for any impact, which gives a risk value of at most 2 with the example matrix. The gate level follows from Table~\ref{tab:matrix} once an impact rating is available. For the eight paths with extended leverage it is H1 for moderate and H2 for major or severe impact. For the other two it is H0 for moderate and H1 for major or severe impact. As an illustration, we assume an impact of at least major, which is plausible where the demonstrated consequence includes vehicle theft, persistent unauthorized code, or the compromise of a trust anchor. Under this assumption the gate requires demonstrated evidence for eight paths and documented evidence for two, whereas the other two approaches give all ten the same very low rating. The baseline inventories alone would have flagged the shared secrets in~\cite{eisenbarth2008keeloq,garcia2016lock,wouters2020dst80} and the interface state in~\cite{garcia2016lock,herrewegen2018bonnet}.

All ten studies report successful attacks. The analysis can therefore expose a false negative of the decision logic, that is, a successfully attacked path for which the gate would not have asked for evidence, and under the stated assumption it finds none. It cannot expose over-classification, because implementations that would have resisted are not in the set. The H0 and H1 outcomes in Section~\ref{sec:example} show that the gate does not ask for demonstrated evidence everywhere, but whether it asks for too much has to be studied on production TARAs.

\section{Discussion}
\subsection{Physical Access as an Exclusion Criterion}
Physical access is a feasibility factor and not a reason to remove a threat. Vehicles are parked unattended, serviced by several organizations, resold, repaired, modified, and dismantled for parts, and a short period of access can have a persistent result such as a recovered key or modified firmware. The method excludes a hardware-assisted path for two reasons only: it is not exposed for any profile in scope, or it is inherently dominated. Both are recorded with their assumptions. Conversely, the method does not inflate hardware risk. Narrow timing windows, characterization per device, destructive preparation, or effects that cannot be repeated raise the factor levels in Step~6 once evidence shows them, and a published attack on a component does not by itself imply high risk: the complete path, the effort of a follower, and the leverage decide.

\subsection{Supplier Interface}
Many vehicle manufacturers do not control the silicon used by a Tier-1 supplier, and the Tier-1 supplier relies on the security subsystem of a semiconductor vendor. The gate turns a TARA result into evidence requests along this chain. An H1 or H2 result translates into questions such as the following. Which secrets are processed outside the security module? Is resistance to SCA or FIA claimed for the primitive in question, and for which configuration? Which debug and lifecycle states are guaranteed in production? What evaluation evidence can be shared? Are keys unique per device? Which reactions to detected faults are fail-secure? SAE J3101-4 provides vocabulary and a list of countermeasures for the SCA and FIA part of this exchange~\cite{saej3101_4}, whereas ISO/SAE 21434 determines why the path matters for the vehicle.

\subsection{Review Criteria}
The following criteria summarize the method for TARA reviews. Each addresses a recurring modeling error.
\begin{enumerate}
    \item Every SCA, FIA, or debug entry names the asset property that is violated; no technique is listed as a threat scenario.
    \item Impact comes from a damage scenario and is not derived from the STRIDE category or from the sophistication of the technique.
    \item The attacker profiles in scope are declared and include the owner and the donor-unit profile.
    \item The key inventory, the decision inventory, and a verified production debug state exist for every ECU in scope.
    \item Simpler competing paths are checked before a specialist SCA or FIA evaluation is requested.
    \item The presence of a countermeasure, such as an HSM, secure boot, a debug lock, or a sensor, is not accepted as evidence of its effectiveness, and neither is its description. A rating that gives credit for resistance refers to E1 or E2 evidence as the level requires.
    \item Each consideration enters once: equipment and time through feasibility, scalability through leverage and the window of opportunity, and consequence through impact.
    \item H0 decisions are recorded as claims with the assumption that reopens them.
\end{enumerate}

\subsection{Limitations}
The method is qualitative. The mapping of Table~\ref{tab:matrix} and the reference values of Table~\ref{tab:sens} are defaults that we derived from the cited attacks. Organizations should calibrate them, and other analysts may assign different factor levels. We have not measured the consistency between analysts. The explicit decision rule is intended to improve it, and a study with independent analyst teams is future work. The application in Section~\ref{sec:example} uses a representative architecture and not a production TARA, and its impact ratings and evidence are assumptions. The retrospective set is purposive and contains only published, successful attacks. It can expose missed cases but not over-classification, and it is subject to hindsight although the gate was considered with $k=0$. Statements about ISO/SAE 21434 refer to the example values of its informative annexes; an organization with different scales has to repeat the analysis of Section~\ref{sec:analysis} for them. The reference values cover non-invasive steps against legitimately manufactured ECUs. Invasive attacks, supply-chain manipulation, sensor spoofing, and electromagnetic interference need their own feasibility interpretation.

\section{Conclusion}
SCA, FIA, and abuse of debug and test interfaces should neither be included in every automotive penetration test nor dismissed because they need physical access. The example feasibility approaches of ISO/SAE 21434 do not support this distinction: the attack-vector-based and CVSS-based approaches rate every physical path as very low, and the attack-potential-based approach lets the rating of one path range over all four levels with the assumed state of knowledge and the attacker profile.

The proposed extension keeps the three techniques at the level of attack methods and adds four elements to the TARA. Attacker profiles make access over the vehicle lifecycle explicit, including the owner and the analyst with a donor unit. Path dominance, defined on the attack-potential factors, orders competing paths and gives the argument of irrelevance through physical access a precise condition. The hardware relevance gate determines from exposure, dominance, impact, leverage, and known attacks whether resistance may be argued, must be documented, or must be demonstrated. The rating rules make the feasibility rating of a hardware step depend on that evidence and give no credit for resistance until the evidence exists.

On the reference architecture, the extension led to treatment or evaluation for five paths that the attack-vector-based and CVSS-based approaches rate at a risk value of 1 or 2, closed one path with supplier evidence, and excluded two with recorded assumptions. Among the ten published attacks that we examined, seven used a debug, boot, diagnostic, or update mechanism and not a cryptographic computation, and eight had extended leverage. This is consistent with verifying debug and lifecycle configuration and inventorying shared secrets first, which costs little compared with a specialist evaluation, and with reserving SCA or FIA evaluation for the cases in which a fleet-wide secret or a trust anchor remains exposed afterwards. Future work is a consistency study with independent analyst teams and the application to a production TARA.

\vspace{0.3cm}
\noindent
\footnotesize{\textbf{Acknowledgment.} The authors used Claude (Anthropic) and ChatGPT (OpenAI) during manuscript preparation for language editing, structural suggestions, brainstorming, and assistance in refining portions of the methodological presentation. All technical content, methodology, literature selection, analysis, results, and conclusions were reviewed and validated by the authors, who take full responsibility for the manuscript.}

\bibliographystyle{IEEEtran}
\bibliography{bibl}

\begin{IEEEbiographynophoto}{Jakub Breier}
received the Ph.D. degree in Applied Informatics from the Slovak University of Technology in Bratislava, Slovakia, in 2013.
He is currently a Senior Cyber Security Manager with TTControl GmbH, Vienna, Austria.
Previously, he held research positions at Silicon Austria Labs and Nanyang Technological University, Singapore.
His research interests include side-channel analysis, fault injection, cryptographic implementation security, embedded systems security, and the physical security of neural network implementations.
\end{IEEEbiographynophoto}
\vspace{-1cm}

\begin{IEEEbiographynophoto}{Xiaolu Hou}
received the B.Sc. and Ph.D. degrees in Mathematical Sciences from Nanyang Technological University, Singapore, in 2012 and 2017, respectively.
She is currently an Associate Professor with the Faculty of Informatics and Information Technologies, Slovak University of Technology in Bratislava, Slovakia.
Her research interests include hardware security of neural network and cryptographic implementations, side-channel analysis, fault injection attacks and countermeasures, and security assessment methodologies.
\end{IEEEbiographynophoto}

\end{document}